# Ultrathin, sputter-deposited, amorphous alloy films of ruthenium and molybdenum

G. Yetik[a], A. Troglia[a], S. Farokhipoor[b], S. van Vliet[a], J. Momand[c], B.J. Kooi[c], R. Bliem[a,d], J.W.M. Frenken[a,d]

[a] *Advanced Research Center for Nanolithography, Science Park 106, 1098 XG Amsterdam, The Netherlands*
[b] *ASML Netherlands B.V., De Run 6501, 5504 DR Veldhoven, The Netherlands*
[c] *Zernike Institute for Advanced Materials, University of Groningen, Nijenborgh 4, 9747 AG Groningen, The Netherlands*
[d] *Institute of Physics, University of Amsterdam, Science Park 904, 1098 XH Amsterdam, The Netherlands*

**Abstract**

Microscopy and diffraction measurements are presented of ultrathin binary alloy films of ruthenium and molybdenum that are obtained by standard sputter deposition. For compositions close to $Ru_{50}Mo_{50}$, we find the films to be amorphous. The amorphicity of the films is accompanied by a significant reduction of the roughness with respect to the roughness of equally thick films of either ruthenium or molybdenum. We ascribe this to the absence of the grain structure that is characteristic of the polycrystalline films of the separate elements.

## 1. Introduction

Thin metal films find broad application in advanced as well as every-day technology as a relatively cost-effective way to customize and improve materials properties, for example to protect against corrosion [1], to lower friction and enhance wear resistance [2-5], as electrically conductive layers [6,7], or, simply, to make a material look metallic. On a microscopic scale, metal films are usually polycrystalline. The strong tendency for metals to crystallize, usually makes these films organize themselves as a conglomerate of small crystallites. Even though most of the atoms in the film can be inside such small crystalline grains, many macroscopic film properties are actually dominated by the grain boundaries between them, where the perfect crystalline stacking is obviously compromised, and by other defects in the crystal structure. For example, the yield strengths measured for polycrystalline films differ strongly from the ideal strengths, predicted by theoretical calculations for perfect single crystals [8], because of the strong effects of dislocations and other crystal defects on the mechanical properties [9]. Chemically, the protective quality of a polycrystalline film can be significantly below that of a single-crystalline one, if atoms or molecules can migrate through the polycrystalline ensemble via the grain boundaries [8]. Even the chemical integrity of the film itself can be at stake, when e.g. oxidation can take place not only at the surface of the film, but also deeply inside, at or via the grain boundaries. In many modern applications, such as low-friction films and optical coatings, extreme, i.e. atomic-scale flatness of thin films would be of advantage [10,11], but the grainy structure of a polycrystalline film also comes with roughness, simply because of the tendency for each grain at the surface of the film to adopt its own equilibrium shape [12]. Rather than connecting into a densely packed, atomically smooth layer, these randomly oriented crystallites typically make the surface look like a cobblestone pavement.

Conceptually, one may recognize two obvious, but extreme solutions to the problems introduced by the polycrystallinity of most metal films: either perfect order, i.e. single crystallinity, or total disorder, i.e. amorphicity. In principle, one should be able to avoid grain boundaries altogether by producing thin, metal films in the form of perfect, single crystals. Unfortunately, even though it is possible to reach lateral grain sizes that can be significantly larger than the thickness of a metal film, it is practically impossible to make the average grain size arbitrarily large and thus completely remove the grain boundaries from thin, crystalline metal films. Moreover, larger grain diameters are accompanied by larger thickness variations of the film – the cobblestone effect. For cases where the lateral grain size would be much larger than the film thickness, this would eventually even lead to places where the film thickness is reduced to zero, at which point the film would be free to break up and expose the underlying material, depending on the surface and interface free energies of the materials involved. The counterintuitive, extreme alternative would be to avoid grain boundaries by going completely in the opposite direction and avoiding crystallinity altogether. For this purpose, the metal would need to be deposited and maintained in a glassy arrangement. This is practically impossible to achieve and maintain for single-element metals.

Interestingly, several metallic alloys are known that can be cast or deposited in an amorphous form. First reports on amorphous metal alloys date back to the early nineteen-sixties [13,14]. Like in the present study, binary metal alloys were used to frustrate crystallization. The rationale behind this approach is that crystallization often requires a significant rearrangement of the internal configuration of the alloy. This can, for example, take the form of segregation of the compound into two different compositions, one enriched in one of the two metals, the other enriched in the other, with at least one of them forming crystallites. This would require migration of atoms over substantially larger distances than the atomic-scale rearrangements required to turn a single-element material into a crystal. Additional kinetic hindrance is introduced when the two constituents have significantly different atomic sizes [15,16]. While a single-element metallic liquid typically would have to be cooled at a rate in the order of $10^{14}$ K/s [17] to reach the glass transition temperature without spontaneous nucleation of crystallites, typical cooling rates required for glass formation of selected binary metal alloys can be lower by more than 11 orders of magnitude [18]. This still calls for dedicated production methods. Since the early work on metallic glasses, a growing variety of new, ever more complex, metallic materials was identified that can be obtained in the form of a glass. One class of examples is formed by high-entropy alloys [19], for which the combination of five or more elements makes that the high configurational entropy of the liquid strongly reduces the melting point and thus reduces the remaining temperature difference that needs to be overcome rapidly to reach the glass transition.

In this article, we concentrate on the classic case of a binary mixture, by investigating alloys of ruthenium and molybdenum. These materials were selected because of their application in reflective optics for lithography with extreme ultraviolet light with a wavelength of 13.5 nm, used in the latest generation of lithography tools in the semiconductor industry [20]. We first demonstrate that sputter deposition of mixtures of these metals with a composition close to $Ru_{50}Mo_{50}$ leads to amorphous thin films with a surprising level of metastability with respect to crystallization. We employ grazing-incidence x-ray diffraction and high-resolution transmission electron microscopy to inspect the amorphicity of these films down to the atomic scale. We further show that these films exhibit an extremely smooth surface when compared to the conventional, polycrystalline surfaces that we obtained by the same sputter deposition procedure for similarly thick layers of the elemental constituents, i.e. Ru or Mo, under the same conditions and on the same substrates.

Thin amorphous films have been demonstrated before for selected binary metal alloys [21], for example in the case of the Cu-Zr system [22]. For the Ru-Mo alloy that is featured in the present article, to our knowledge, no direct observations are available for the amorphicity of thin films, albeit measurements in [23] on the electrical resistivity of electron-beam deposited RuMo films were interpreted already in 1978 as an indication for their amorphous structure in a certain range of compositions.

That the surface roughness of thin amorphous metal alloy films can be very low [21,24] and that it increases upon the change from amorphous to polycrystalline, has been found before, for example in annealing experiments [25] on ZrAlMoCu metallic glass films and in the comparison of polycrystalline Zr and Cu films and amorphous CuZr alloy films [22]. Similar indications were reported for ZrCuAlNi metallic glass films [26]. For the CuZr-case, the larger roughness of the polycrystalline films was ascribed to their columnar structure [22].

A special feature of the binary phase diagram of Ru-Mo mixtures is the existence of an entropy-stabilized configuration for a narrow range of compositions around $Ru_3Mo_5$, the so-called σ-phase [27-30], a tetragonal lattice (space group $P4_2/mnm$) with five inequivalent sublattices of atomic sites. This phase occurs at temperatures below the eutectic point, i.e. the lowest melting point for any of the mixtures. Rather than to cool the mixture from the melt, we concentrated on ultrathin films that we routinely deposit on Si(100) wafers, terminated by a native oxide layer, by simultaneous sputter deposition of Ru and Mo from separate sources.

## 2. Experimental

*2.1. Film preparation*

A Polyteknik Flextura M506 S [31] system was used for the sputter deposition [32] of all thin films in this work. In contrast to Ru-Mo growth studies reported in the literature [23,33], our substrates were kept at room temperature. All films were deposited on p-doped Si(100) substrates; the native oxide on these substrates was not removed prior to deposition. For the production of thin $Mo_xRu_{100-x}$ alloy films, Mo and Ru were co-deposited from separate sources by RF and DC sputtering, respectively. We used different types of power supplies for the two metals for practical reasons, but we do not expect this to influence the results significantly. The base pressure of the sputter deposition system was $1 \times 10^{-7}$ mbar and the argon (Ar) pressure was $2 \times 10^{-3}$ mbar or higher during the deposition. In addition to the alloy films, also regular films of pure molybdenum and pure ruthenium were deposited by DC sputtering. During deposition, the substrates were rotated, in order to optimize the homogeneity of the deposition.

Table 1. provides typical examples of the sputter-deposition conditions employed for the production of the alloy films discussed in this article. The first column shows the resulting compositions that were derived from the energy-dispersive x-ray spectroscopy (EDX) measurements in our scanning electron microscope (SEM, see below).

| Alloy composition (EDX) | Power level and type of power supply for Ru (W) | Power level and type of power supply for Mo (W) | Ar pressure (mbar) | Deposition rate (nm/s) |
|---|---|---|---|---|
| Pure Ru | 200 DC | – | $2 \times 10^{-3}$ | 0.30 |
| $Ru_{77}Mo_{23}$ | 75 DC | 100 RF | $2 \times 10^{-3}$ | 0.16 |
| $Ru_{43}Mo_{57}$ | 75 DC | 250 RF | $2 \times 10^{-3}$ | 0.28 |
| Pure Mo | – | 75 DC | $1.33 \times 10^{-2}$ | 0.15 |

**Table 1.** Alloy film compositions, measured by SEM-EDX, in combination with the employed deposition powers, the working gas pressures used, and the types of power supplies (DC versus RF) used in the sputter deposition process. The deposition rates in the final column were derived from the measured film thicknesses.

In this way, films were obtained with thicknesses ranging from 4 to 30 nm and compositions ranging from pure Ru to pure Mo.

*2.2. Characterization*

Measurements of the thicknesses of the deposited metal films were carried out in the following way. Prior to metal deposition, which was usually done simultaneously on multiple substrates, one of these substrates was partly coated by a photoresist film. After the metal deposition, the resist was washed off by acetone and isopropanol and this cleaning procedure also removed the deposited metal film from those parts of the substrate that had first been coated by the resist, while the metal film on other parts of the substrate stayed in place and remained unaffected. This procedure enabled us to accurately measure the height difference between the bare regions where the resist was removed and the regions that had not been coated by the resist and that were covered by the thin metal film. These height difference measurements were done by use of a profilometer and by atomic force microscopy (AFM). This provided us with accurate measurements of the film thicknesses. Samples used to measure the film thickness were not utilized for further experiments to avoid possible contamination from the employed chemicals and their residues. The films used in this article had thicknesses ranging from 4 to 30 nm.

We employed three different imaging techniques to inspect the metal films and the substrates. The first of these was tapping-mode AFM [34], which we used to image the surface and measure the topography, the film thickness (see above) and the surface roughness. For this, we used a Bruker Dimension Icon AFM [35] system in ScanAsyst-air mode with silicon tips on silicon nitride cantilevers of Bruker's SCANASYST-AIR or SCANASYST-AIR HR type. AFM image processing was done by using the NanoScope Analysis version 2.00 software. The surface roughness, defined as the root-mean-square height variation, was obtained from the asymptotic value of the height correlation function, measured in each AFM image along the fast scanning direction.

In addition, we employed scanning electron microscopy (SEM), for which we made use of an FEI Verios 460 SEM system [36], with a Schottky field electron gun. The SEM images were taken at an electron energy of 5 keV and a beam current of 100 pA, unless indicated differently, and the samples were placed at approximately 4 mm distance from the analyzer. For all SEM images, we used the immersion field mode, in order to optimize the spatial resolution [37].

For transmission electron microscopy (TEM) [38], cross sectional specimens were prepared with an FEI Helios G4 CX dual beam system at 30 kV ion energy and polished at 5 keV and 2 keV to remove residual surface damage. These specimens were analyzed with a double-aberration-corrected FEI Themis Z scanning transmission electron microscopy (STEM) system at 300 kV. High-angle annular dark-field (HAADF) STEM images were recorded with a ~200 pA probe current, at a convergence semi-angle of 21 mrad and with HAADF collection angles of 61–200 mrad. In order to avoid spurious edge effects in the two-dimensional fast Fourier transforms, a Hann filter was used to suppress the contribution of the image edges to the FFT patterns.

Additional structural information came from standard, grazing-incidence X-ray diffraction (GI-XRD) measurements that were performed with Cu-k$\alpha$ radiation (wavelength of 0.15406 nm), using a Panalytical X'Pert MPD diffractometer. The incidence angle was set at a constant, low value of 0.5° with respect to the sample's surface plane, for all measurements.

We used two techniques to measure the composition of the deposited alloy films. The first of these was energy-dispersive x-ray spectroscopy (EDX) [39]. Within the SEM system, we performed EDX measurements with the use of an Oxford Xmax 80 detector. Typically, the SEM-EDX measurements were performed on multiple locations on the sample. EDX spectrum imaging in the TEM system was performed with a probe current of 1 nA, where the spectra were recorded with a Dual-X system, providing in total 1.76 sr EDX detector.

The second technique that we employed to measure the surface composition of the thin films was X-ray photoelectron spectroscopy (XPS) [40]. Measurements with this technique were performed *ex-situ* with X-rays emitted from a monochromatic Al-K$_\alpha$ (1486.6 eV) source operating in an ultrahigh vacuum setup (base pressure better than $1.0 \times 10^{-9}$ mbar) equipped with a Scienta Omicron R4000 HiPP-3 analyzer (swift acceleration mode, 1 mm slit entrance) for elemental and chemical characterization of the surfaces of the films. Measured XPS peak shapes and intensities were fitted using the software KolXPD from Kolibrik [41].

## 3. Results

*3.1. Observation of amorphous alloy structure*

Figure 1 concentrates on the structural characteristics of approximately 21 nm thick RuMo alloy films with two different compositions. The $Ru_{43}Mo_{57}$ film is closest to the eutectic composition of $Ru_{42}Mo_{58}$ [28,30]. A strong fingerprint of the amorphous nature of this film is provided by the upper GIXRD pattern in Fig. 1a. It contains two very broad peaks, similar to the diffraction pattern from a liquid [42,43]. The complete absence of sharp peaks indicates that if some of the Ru and/or Mo atoms would still have formed crystals, their fraction of the total cannot establish more than approximately 1% of the total number of atoms, based on the sensitivity and the noise level in the data. Note that the sharp peaks, labeled by an asterisk, stem from the Si(100) substrate and have low intensities due to the grazing-incidence geometry of the experiment that was chosen to maximize the signal from the film. The sharpness of these Si peaks testifies to the angular resolution of the diffraction measurement. In order to be certain that the broad GIXRD peaks of the alloy correspond to an amorphous configuration, rather than a nano-crystalline arrangement, for which the small crystallite size might perhaps lead to similarly broad peaks, we inspected the $Ru_{43}Mo_{57}$ film also with high-resolution cross-sectional TEM. The result is shown in Fig. 1b. While the Si(100) substrate exhibits its familiar, well-ordered lattice structure, the alloy film is completely disordered on all length scales, down to atomic dimensions, as can be verified directly in the enlarged section in Fig. 1c. The spatial information in the TEM image is consistent with the GIXRD spectrum, as can be recognized from the two-dimensional fast Fourier transform (2D FFT) in Fig. 1d that was taken from the TEM region in Fig. 1c, completely within the alloy film. As expected from the qualitative impression of the TEM image itself, the 2D FFT contains no orientational preferences. The broad ring reflects the short-range correlation between the atoms in the amorphous film and their direct neighbors, and its radius coincides with the position of the main peak in the GIXRD pattern. Interestingly, the surface of the film makes a smoother impression in the TEM image than the surface of the thin native oxide layer on which the film was deposited. We will return to this point later.

The lower curve in Fig. 1a shows the GIXRD spectrum measured for an equally thin film of the more ruthenium-rich alloy, $Ru_{77}Mo_{23}$. It is dramatically different from the diffraction pattern from the amorphous alloy, $Ru_{43}Mo_{57}$, and shows a rich collection of sharp peaks that can all be associated with a polycrystalline layer with the hexagonal close-packed structure of ruthenium. Interestingly, the peak positions correspond to lattice constants of a = 0.272 nm and c = 0.432 nm, which indicates an expansion of the volume of the lattice unit cell by 2.0 % ± 0.35 % with respect to the regular lattice of ruthenium. This indicates that these diffraction peaks do not come from pure ruthenium, but from a solid solution of the somewhat larger molybdenum atoms in ruthenium[22]. The lattice expansion can be associated with a composition of $Ru_{84}Mo_{16}$, with an error margin of no more than ± 3 % on the two atomic concentrations. The widths of the diffraction peaks exceed the resolution and can be used to estimate an average grain size of approximately 10 nm, a significant fraction of the film thickness. We speculate that the ruthenium-rich crystals that the GIXRD spectrum is indicative of are formed out of the deposited, 77:23-mixture by segregation. The remaining material in the film must then be enriched in molybdenum. One might expect this remaining component to form body-centered cubic crystallites, i.e. with the lattice structure of molybdenum (and with a slightly reduced lattice constant due to the dissolved ruthenium). However, the GIXRD pattern does not contain any additional peaks. This means that the molybdenum-enriched component must be present in a different form, which does not generate distinct, i.e. sharp features in the GIXRD. In order to further elucidate the segregated structure for this composition, additional experiments will be required, which we will report in a future publication.

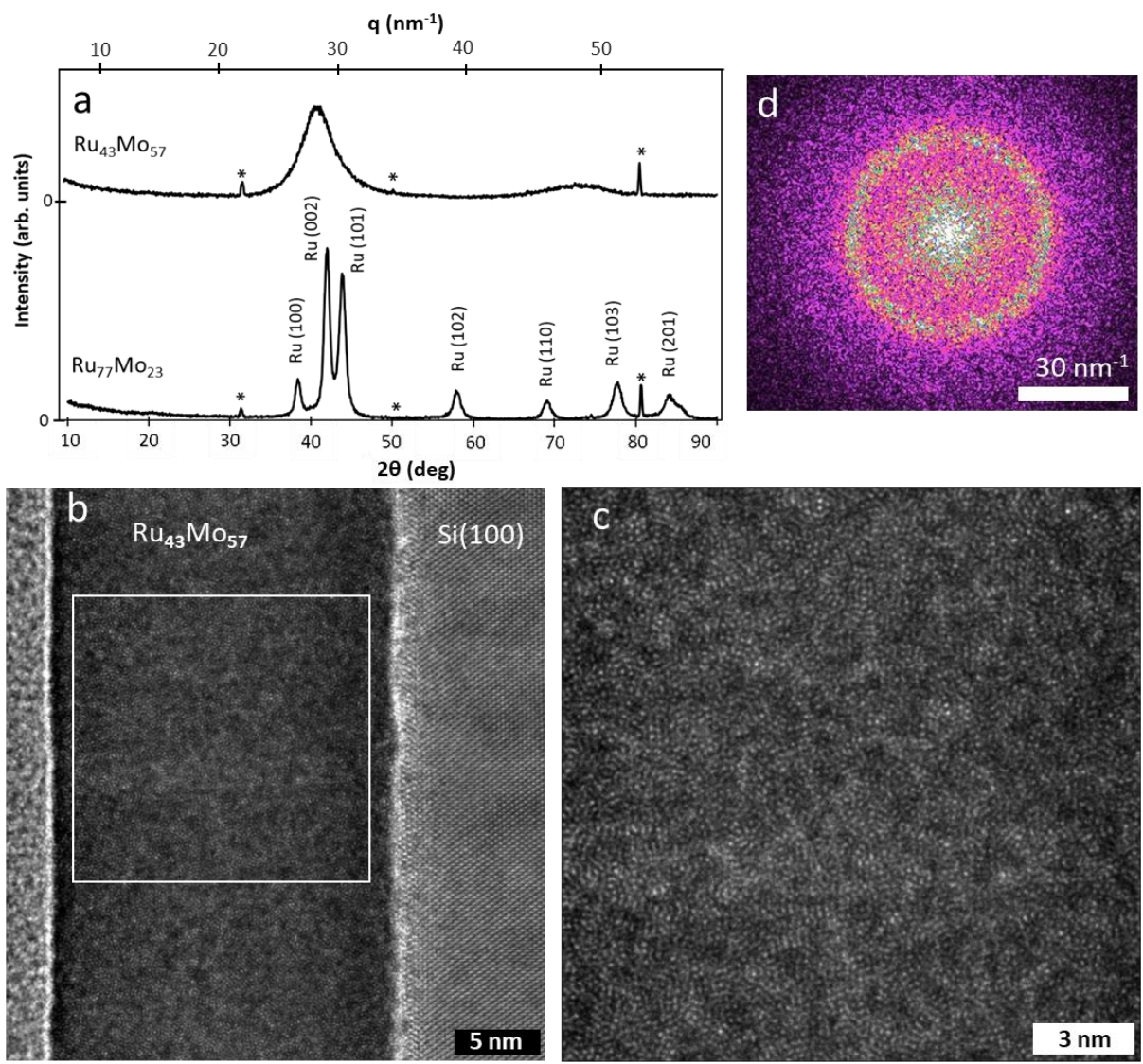

**Fig. 1.** (a) GI-XRD measurements from 21 nm thick $Ru_{43}Mo_{57}$ (top) and $Ru_{77}Mo_{23}$ (bottom) alloy films. Along the upper horizontal axis of the panel, the scattering angles are converted into the corresponding reciprocal-space scale. The sharp peaks indicated by asterisks are artifacts originating from the Si(100) substrate. (b) Cross-sectional high-resolution TEM micrograph of the $Ru_{43}Mo_{57}$ alloy film and the Si(100) substrate. Note the crystalline structure of the Si substrate, the structural disorder of the thin native oxide layer by which the substrate is terminated and the complete absence of crystalline order in the metal alloy film on top of that. The white rectangle indicates the region that is displayed on a magnified scale in panel (c) to emphasize the disordered arrangement in the alloy. Panel (d) represents the two-dimensional fast Fourier transform of this region, which matches the features in the upper GIXRD curve of panel (a), measured from the same composition.

The diffraction pattern and TEM observations of the amorphous nature of the $Ru_{43}Mo_{57}$ film were characteristic for the entire film, rather than specially selected 'best cases'. Figure 2a shows the same combination of a selected region from a cross-sectional high-resolution TEM micrograph of the $Ru_{43}Mo_{57}$ alloy film and its two-dimensional fast Fourier transform that was shown already in Figs. 1.c,d. For completeness, we also show Figs. 2b and c of two other, equally large regions from the same film and the corresponding 2D-FFT patterns. The arrows point to what one might interpret as weak indications of local order, but that does not stand out as ordered domains in the corresponding real-space images. Apart from such occasional local intensities, all two-dimensional Fourier transforms were fully consistent with the smooth pattern for Fig. 2a.

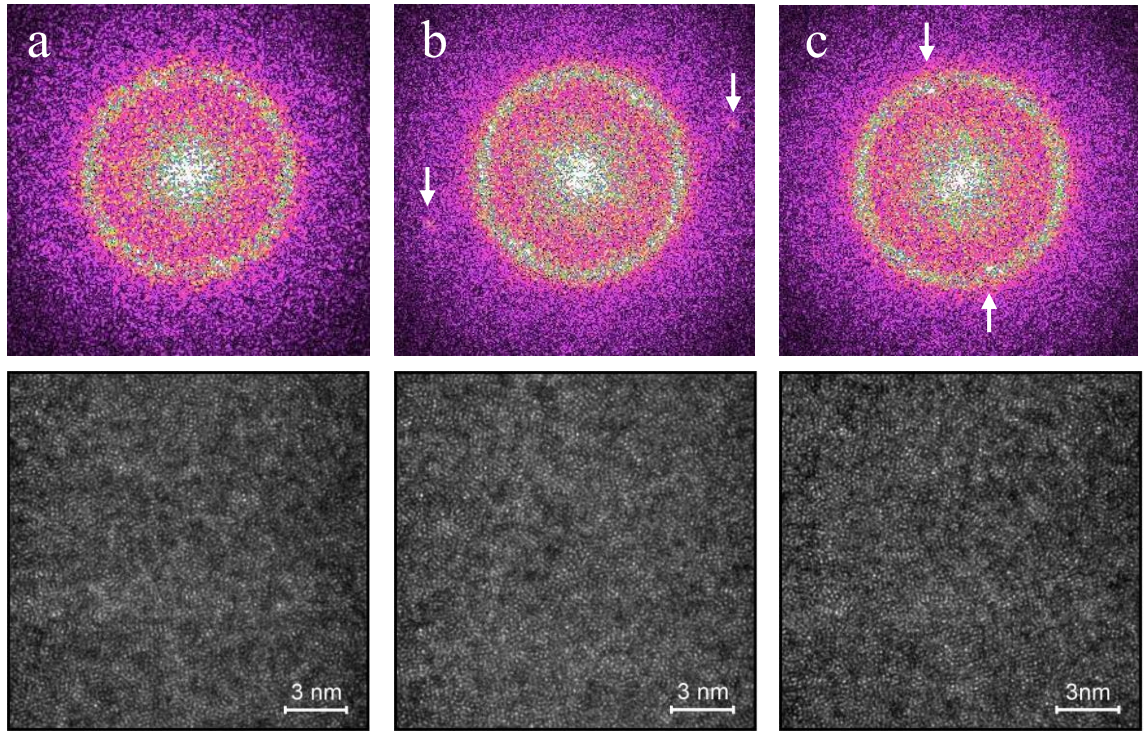

**Fig. 2.** Three selected regions from cross-sectional high-resolution TEM micrographs of the $Ru_{43}Mo_{57}$ alloy film, each taken completely within the alloy film. The corresponding two-dimensional fast Fourier transforms all display the same broad ring, typical for an amorphous arrangement. The arrows in panels (b) and (c) indicate weak, additional signatures of local order that we have observed in some of these local 2D FFTs, but that we cannot recognize readily in the corresponding TEM images.

*3.2. Film composition*

Figure 3 shows the XPS spectra of a polycrystalline $Ru_{77}Mo_{23}$ thin film (top) and an amorphous $Ru_{43}Mo_{57}$ thin film (bottom). The survey spectra (left) have been acquired at 500 eV pass energy, while the detailed Ru 3d and Mo 3d regions (right) have been acquired at 300 eV pass energy. The survey spectra only contain peaks associated with ruthenium, molybdenum and oxygen.

The detailed Ru 3d and Mo 3d regions have been fitted using Voigt functions, taking into account the presence of metal oxides and the broadening due to the Coster-Kronig effect. A Shirley function has been used for background correction.

The surface fractions of ruthenium and molybdenum were inferred from the total areas of the Ru 3d and Mo 3d peaks, obtained from the fits and the corresponding photoionization cross sections, according to the general equation:

$$x_i = \frac{\frac{A_i}{\sigma_i}}{\sum_{j=1}^{N}\left(\frac{A_j}{\sigma_j}\right)},$$

where $A$ is the area of a peak, $\sigma$ is the corresponding photoionization cross-section, $i$ and $j$ refer to the elements ruthenium and molybdenum, and $N$ stands for the total number of elements, in this case 2. The surface compositions that we obtained in this way for the two alloys in Fig. 3, are given in the two panels on the left, with the overview spectra. Within the error margins, the measured surface compositions of $Ru_{79}Mo_{21}$ and $Ru_{43}Mo_{57}$, are equal to the (average) bulk compositions of $Ru_{77}Mo_{23}$ and $Ru_{43}Mo_{57}$ that we obtained from SEM-EDX measurements on the same samples (see Table 1).

The oxygen peaks are expected due to the exposure of the surface to air after the deposition, leading to surface oxidation and water adsorption. The XPS spectra contain a minor molybdenum peak at 235 eV, stemming from molybdenum oxide. Note that the surface of the $Ru_{43}Mo_{57}$ alloy contains more oxygen than that of the $Ru_{77}Mo_{23}$ alloy, which may be attributed to the stronger oxidation resistance of ruthenium capping layers [44,45].

A possible C 1s peak, due to carbon contamination during deposition, would be located around 284.5 eV and would thus overlap with the Ru 3d peak structure. When fitting the Ru 3d peak, we arrive at the correct 3:2 peak ratio, from which we conclude that carbon contamination can be neglected.

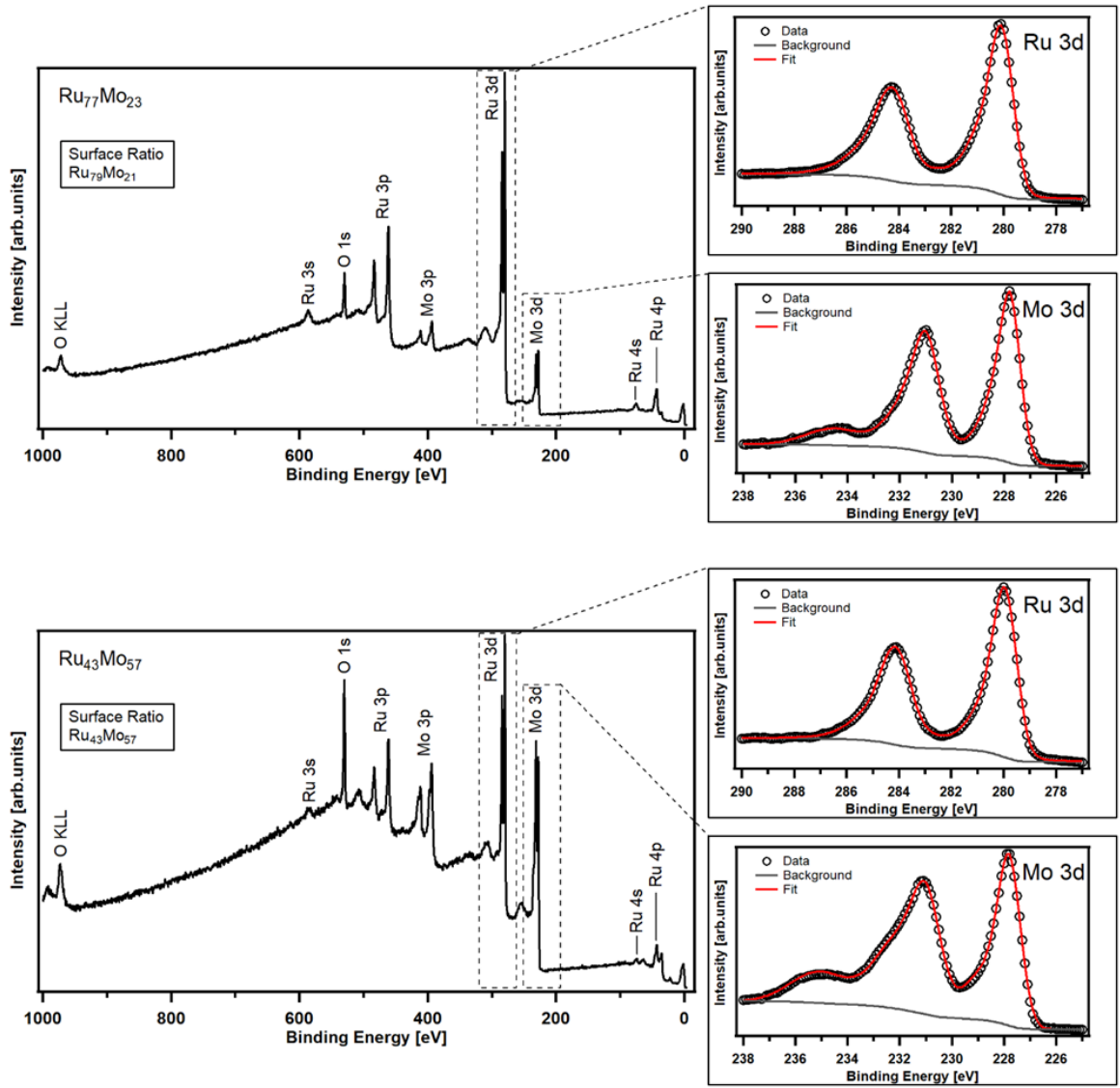

**Fig. 3.** XPS spectra for a 20 nm thick, polycrystalline film with a composition of $Ru_{77}Mo_{23}$ (top) and for a 21 nm thick, amorphous film with a composition of $Ru_{43}Mo_{57}$ (bottom). The panels on the right show the detailed spectra for the two thin films around the Ru 3d and the Mo 3d regions, and the corresponding fits (red curves).

The bottom right panel of Fig. 4 repeats the TEM image of Fig. 1.b. The other panels show the simultaneously acquired EDX maps for four elements, ruthenium, molybdenum, silicon and oxygen. As the ruthenium and molybdenum maps clearly show and as is quantified in the concentration curves for the two metallic elements in the bottom left panel, the concentrations of the two metals go through a total of close to five full cycles of approximately 15% variation, with the two metals varying in antiphase with each other. This variation is a direct consequence of the source geometry of our sputter deposition chamber in combination with the continuous sample rotation during deposition. As explained in the main article, we produce our alloy films by co-deposition from separate, pure molybdenum and pure ruthenium targets. These sputter targets are arranged off-axis with respect to the rotation axis of the sample platform, each directed towards the center of the platform. This platform carries multiple samples, with each sample placed at a certain distance from the rotation axis. During the deposition, the sample platform rotated at a speed of 4 revolutions per minute (RPM), periodically bringing each sample closer to the molybdenum target and further away from the ruthenium target and vice versa. The period of the observed concentration variations matches this rotation. Interestingly, this variation implies that the results in Fig. 4 and in the rest of this article are all for alloy films that internally contain a range of compositions. The entire, 21 nm thick film of Fig. 4 can be seen to be amorphous, as is also confirmed by the GIXRD measurements, even though the composition within the film varies between $Ru_{35}Mo_{65}$ and $Ru_{50}Mo_{50}$. This shows that the alloy is amorphous over at least this range of compositions.

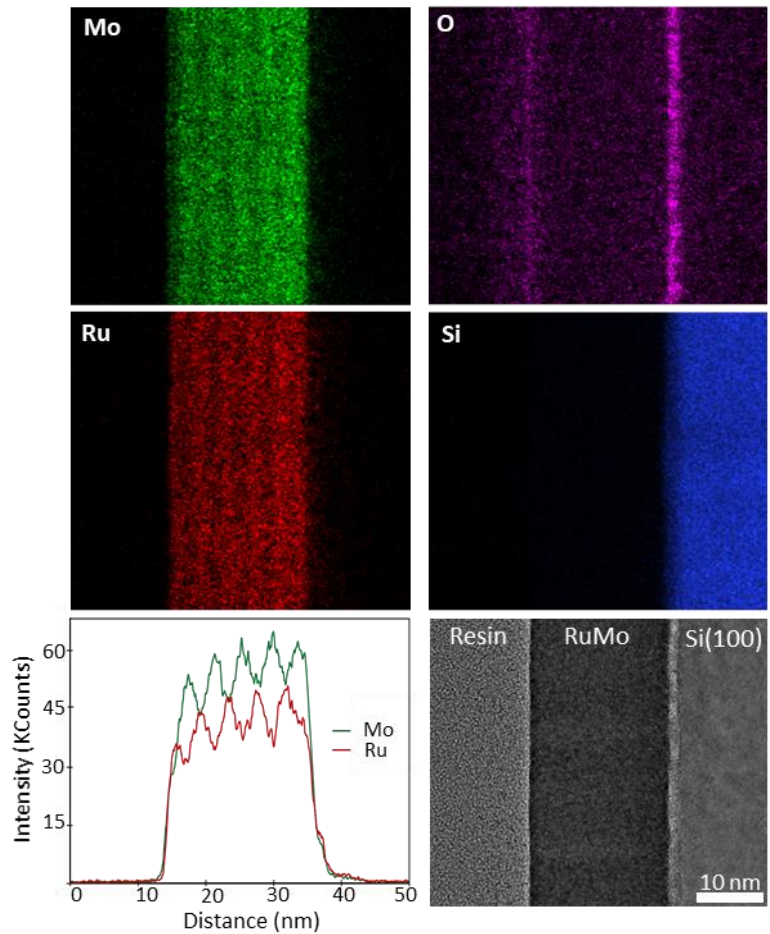

**Fig. 4** HAADF-STEM image (bottom right panel) and simultaneously measured EDX maps on a 21 nm thick film of the $Ru_{43}Mo_{57}$ alloy (top four panels) for four elements: Ru (red), Mo (green), Si (blue) and O (purple). The corresponding concentration curves for Ru and Mo (bottom left panel) demonstrate a periodic 15% variation in the composition, with the Mo and Ru concentrations changing in antiphase with one another. This variation can also be recognized directly as the striations in the EDX panels of Mo and Ru.

*3.3. Low surface roughness on amorphous alloy film*

SEM images show a fully structureless surface on the $Ru_{43}Mo_{57}$ films, in full contrast with the familiar network of grains and grain boundaries that we observe in SEM images on thin films of pure ruthenium or pure molybdenum. In fact, focusing the electron microscope on these alloy films was only possible by virtue of special features, such as the edges of the films, as the surfaces of the films did not offer any discernable contrast. In order to obtain more quantitative information on the height variations, we inspected the surfaces of the films with AFM. The upper left panel of Fig. 5 shows a large-area scan of a 30 nm thick $Ru_{43}Mo_{57}$ film. In spite of the large scan range, all height variations are fully captured by the tight ± 0.7 nm height range of the color bar. The enlarged region shown in the upper right panel of Fig. 5 indicates that there are mild undulations in the height, with a typical lateral length scale in the order of 10 nm. The surface roughness of the alloy, that can be derived from these images as the root-mean-square (rms) height variation around the average surface plane, is spectacularly low, 0.26 nm. Within the error margins, the alloy film seems not to introduce any additional roughness with respect to the roughness of the native oxide on the Si(100) substrate, on which the metal films were deposited.

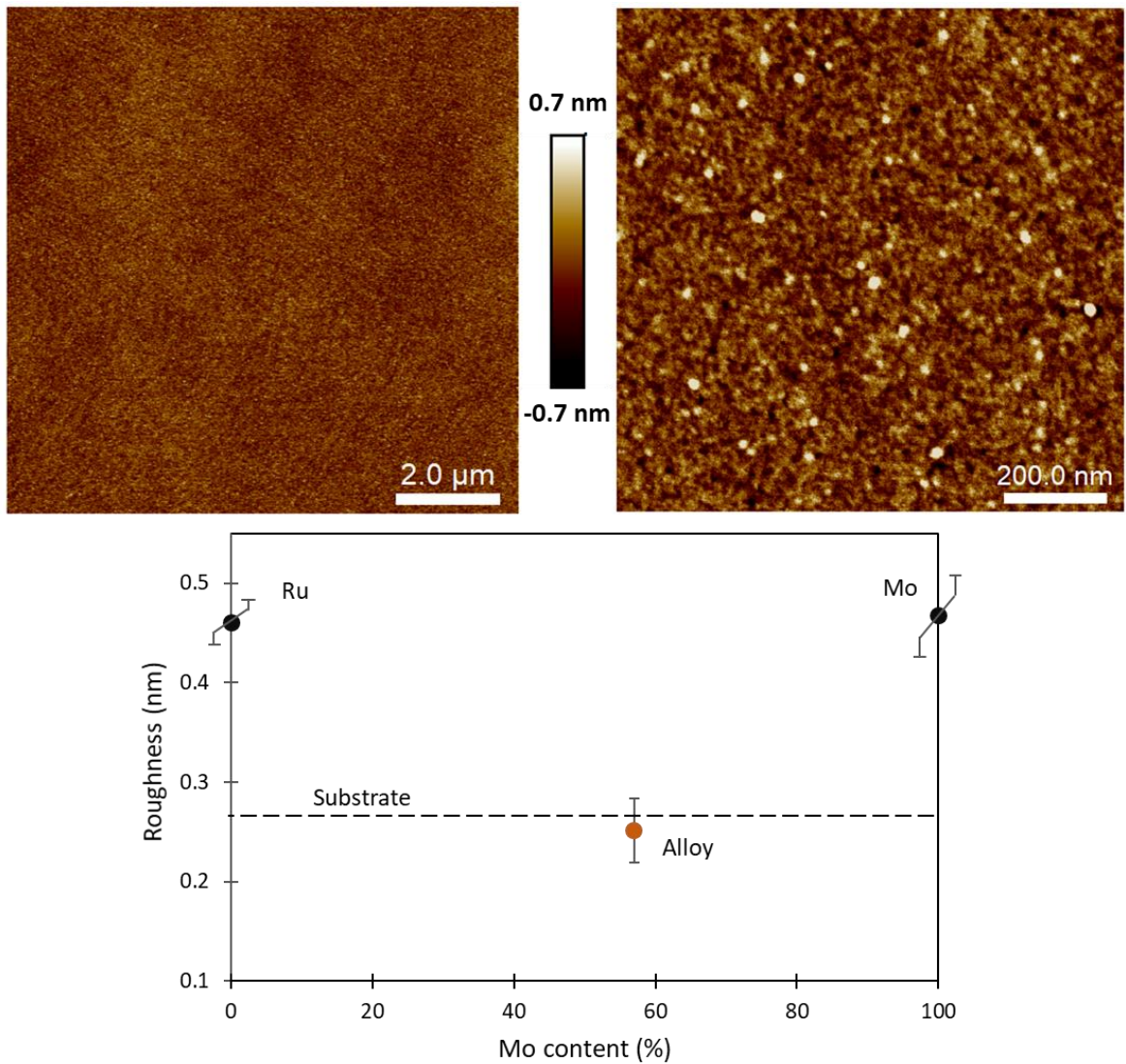

**Fig. 5.** The two upper panels are AFM height micrographs obtained on a 30 nm thick film of the $Ru_{43}Mo_{57}$ alloy, with image sizes of 10 μm × 10 μm (left) and 1 μm × 1 μm (right). Both images are displayed with the same conversion of height variations into colors, as indicated by the color bar. The pattern of parallel lines, vaguely visible in the larger-scale image, are an artifact resulting from a minor mechanical vibration. The bottom panel compares the values of the surface roughness (rms height variation) for 20 nm thick films of pure Ru, pure Mo and the $Ru_{43}Mo_{57}$ alloy, measured from AFM images with sizes of 1 μm x 1 μm. The lower dashed line indicates the surface roughness encountered on the substrate, prior to deposition.

The roughness of the metal films, including the RuMo alloy, is plotted in the bottom panel of Fig. 5. Each point in the graph is the result of several AFM measurements on a different film. Whereas one might have anticipated the $Ru_{43}Mo_{57}$ mixture to adopt at best some sort of average of the roughness values of pure ruthenium and pure molybdenum films, and that any tendency to segregate would introduce extra roughness, the data clearly shows that the mixed film is significantly smoother than pure ruthenium and pure molybdenum. In fact, the alloy does not add any roughness on top of that of the underlying substrate; within the error margin it seems to be even slightly smoother than the substrate. Note, that the alloy film in Fig. 5 is actually thicker than the films of the two pure metals. We investigated a small number of films with a larger thickness, for which we find the same trend. For example, a 100 nm film of $Ru_{43}Mo_{57}$ exhibited a roughness of 0.43 nm, whereas the roughness of an equally thick, pure Ru film was 0.75 nm.

## 4. Discussion and conclusion

There is a simple scenario, sketched in Fig. 6, that illustrates how the microstructure of a film, in particular the film's polycrystallinity or amorphicity, plays a dominant role in the roughness of the film. We start by emphasizing that our observations for pure ruthenium and pure molybdenum films are quite typical. The SEM micrographs and the AFM images for the pure metals show, in both cases, a polycrystalline arrangement with a significant roughness that is introduced by the grainy morphology of these films, with its typical network of grain boundary grooves (left panel of Fig. 6). The increase of the roughness with increasing film thickness for the pure films is a direct consequence of the increasing lateral size of the grains, which is accompanied by increasing height variations [12]. We have also performed a few additional measurements for thicker films, up to film thicknesses of 100 nm. The results from those are consistent with what we report in the article for the thinner films.

These grain-related aspects are all absent for films with the $Ru_{43}Mo_{57}$ composition (right panel of Fig. 6), for which both the GIXRD and the TEM data provide atomic-scale evidence that they are amorphous. The atomic arrangement of such an amorphous film does not suffer from crystal defects such as the familiar grain boundaries, where otherwise the structure would be compromised and grooves would develop at the film surface, in order to minimize grain boundary energy costs. In fact, for an amorphous film, the opposite might take place. The minimization of the surface free energy by virtue of the transient atomic mobility during the (sputter) deposition process may rather lead to surface *smoothening* instead of roughening [24]. That such smoothening is indeed at play, seems to be suggested by the reduction in apparent roughness that the surface of the $Ru_{43}Mo_{57}$ alloy film in TEM micrographs in Fig.1 and Fig.4. exhibits with respect to its interface with the underlying substrate. This also seems to be confirmed by the slightly lower AFM-value of the surface roughness of films with that composition with respect to that of the substrate (Fig. 5), albeit that this modest difference remains within the statistical error margin. The observed roughness lies in the range of the height variations that one otherwise only finds in the case of (near)-ideal layer-by-layer growth or in the case of (near)-ideal step-flow growth of single crystals. In the layer-by-layer case, usually, the roughness progressively increases, which typically washes out the layer-by-layer character within the first ten atomic layers, i.e. the first few nanometers. Also, in step-flow growth, statistical variations in the deposition gradually roughen up the surface. The smoothness of the $Ru_{43}Mo_{57}$ alloy film is superior to that of its crystalline counterparts and greatly exceeds our expectations.

A few extra words are in place on the precise composition, for which we find our films to be amorphous. The TEM image of Fig. 1 was obtained for a film with an average composition of $Ru_{43}Mo_{57}$. However, the spatially resolved EDX observations on these cross sections that we introduced above, showed that the composition varied as a function of depth. The 21 nm film contained a total of 4.9 cycles of a compositional variation that ranged between $Ru_{35}Mo_{65}$ and $Ru_{50}Mo_{50}$. Careful inspection of the TEM images did not reveal crystallization over this entire range of compositions, which suggests that amorphous layers can be deposited over a relatively wide range of compositions, similar to what was found for the Cu-Zr system [22].

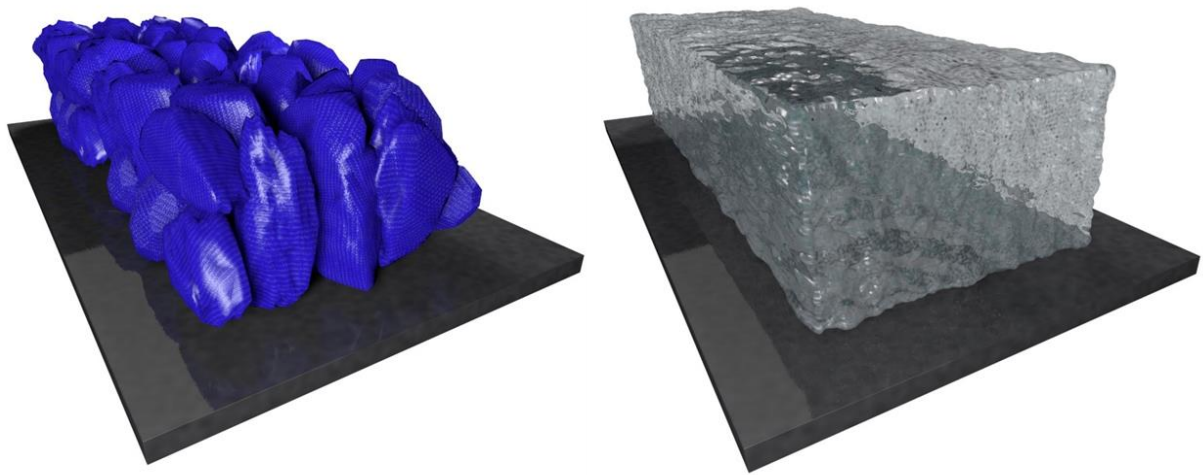

**Fig. 6.** Schematic of the two configurations encountered in this study. (Left) The typical structure of a metal film is polycrystalline, characterized by a collection of crystalline grains that, together, lead to a rough surface. (Right) The complete absence of grains makes an amorphous alloy film extremely smooth [46].

These results have implications that can be of relevance for potential applications of Ru and Mo coatings, for example in the context of EUV lithography technology. As we mentioned already, many problematic properties of thin metal films that limit their performance in practice, stem from their grainy nature. These disadvantages are all reduced or removed when the film can be made amorphous, providing improvements in terms of smoothness, mechanical strength, impermeability, friction and wear, corrosion resistance, and other properties. What our results for alloys of ruthenium and molybdenum show, is that with the industrially familiar technique of sputter deposition, it is relatively straightforward to deposit amorphous alloy films of molybdenum and ruthenium. The amorphicity seems to be rather robust with respect to variations in the precise alloy composition, at least for this particular combination of metals. From the perspective of applications, this even provides some flexibility to optimize the film composition for specific, chemical, optical, or mechanical behavior, while maintaining the amorphous structure and the accompanying advantages.

**CRediT authorship contribution statement**

**Görsel Yetik:** Conceptualization, Sample preparation, AFM and SEM Analysis and Interpretation, Validation, Writing – original draft, review & editing. **Alessandro Troglia:** Sample preparation, XPS Analysis and Interpretation, Writing – review & editing. **Saeedeh Farokhipoor:** GIXRD Analysis and Interpretation, Writing – review & editing. **Stefan van Vliet:** AFM Analysis, Writing – review & editing. **Jamo Momand:** TEM Analysis and Interpretation, Writing – review & editing. **Bart J. Kooi:** TEM Analysis and Interpretation, Writing – review & editing. **Roland Bliem:** Conceptualization, Supervision, Writing – review & editing. **Joost W.M. Frenken:** Conceptualization, Supervision, Interpretation, Writing – original draft, review & editing.

**Declaration of Competing Interest**

The authors declare that they have no known competing financial interests or personal relationships that could have appeared to influence the work reported in this paper.

**Acknowledgments**